\documentclass[a4paper,11pt]{article}

\usepackage{contribution}
\usepackage{subfig}

\newcommand{\weblink}[2][]{%
    \ifthenelse{\equal{#1}{}}%
    {\textnormal{\url{#2}}}%
    {\textnormal{\href{#2}{#1}}}%
}

\newcommand{\acknowledgements}[1]{%
  \bigskip\bigskip
  \textsf{\textbf{\Large Acknowledgements}} \\[2ex]
  {#1}
  \bigskip
}

\def\beq{\begin{equation}}
\def\eeq#1{\label{#1}\end{equation}}
\def\eeqn{\end{equation}}

\def\beqa{\begin{eqnarray}}
\def\eeqa#1{\label{#1}\end{eqnarray}}
\def\eeqan{\end{eqnarray}}

\let\bar=\overbar

\def\etal{{\it et al.}}

\def\Dslash{\not{\hbox{\kern-4pt $D$}}}
\def\dslash{\not{\hbox{\kern-2pt $\del$}}}

\def\msb{{\bar{\ssstyle M \kern -1pt S}}}

\newcommand{\contribution}[7][]{%
  \clearpage
  \thispagestyle{plain}
  \ifthenelse{\equal{#1}{}}
  {\hypersetup{pdftitle={#2}}}
  {\hypersetup{pdftitle={#1}}}
  \hypersetup{pdfauthor={{#3} {#4}}}
  {\centering\normalfont\LARGE\bfseries\sffamily #2 \par\nobreak}
  \lhead{}
  \chead{%
    \textit{\footnotesize XIV International Conference on Hadron Spectroscopy
      (\weblink[\textit{hadron2011}]{http://www.hadron2011.de}), 13-17 June 2011, Munich, Germany}%
  }
  \rhead{}
  \bigskip
  \begin{center}
    {#3} {#4}\ifthenelse{\equal{#6}{}}{}{\footnote{\weblink[#6]{mailto:#6}}}
    \ifthenelse{\equal{#7}{}}{}{#7} \\
    \textit{#5}
  \end{center}
  \bigskip
}

\renewcommand{\abstract}[1]{%
  \begin{center}
    \begin{minipage}{0.85\textwidth}
      \begin{footnotesize}
        #1
      \end{footnotesize}
    \end{minipage}
  \end{center}
  \bigskip
}

\begin{document}

% % % % % % % % % % % % % % % % % % % % % % % % % % % % % % % % % % % % % % % % %
% your proceedings
%%%%%%%%%%%%%%%%%%%%%%%%%%%%%%%%%%%%%%%%%%%%%%%%%%%%%%%%%%%%%%%%%%%%%%%%%%%%%%%%%
%
% template for hadron2011 contribution
%
% please do not rename this file
%
% to create document run
%
%     pdflatex hadron2011.tex
%
%%%%%%%%%%%%%%%%%%%%%%%%%%%%%%%%%%%%%%%%%%%%%%%%%%%%%%%%%%%%%%%%%%%%%%%%%%%%%%%%%
{  % do not remove

%%%%%%%%%%%%%%%%%%%%%%%%%%%%%%%%%%%%%%%%%%%%%%%%%%%%%%%%%%%%%%%%%%%%%%%%%%%%%%%%%
% please define your macros here

%
%%%%%%%%%%%%%%%%%%%%%%%%%%%%%%%%%%%%%%%%%%%%%%%%%%%%%%%%%%%%%%%%%%%%%%%%%%%%%%%%%

%%%%%%%%%%%%%%%%%%%%%%%%%%%%%%%%%%%%%%%%%%%%%%%%%%%%%%%%%%%%%%%%%%%%%%%%%%%%%%%%%
% define title, author, and address
% contribution[short title]{title}{author first name}{author last name}{author address}{author email}{collaboration}
% the short title will appear in the page headers and the TOC of the book of proceedings
% the last two arguments may be left empty
\contribution[Single and Double Pion Photoproduction off the Deuteron]  % short title (optional)
{Single and Double Pion Photoproduction \\ off the Deuteron}  % title
{Manuel}{Dieterle}  % first and last name of author
{Department of Physics \\
  University of Basel \\
  CH-4056 Basel, SWITZERLAND}  % author address
{manuel.dieterle@unibas.ch}  % author email optional
{for the A2 Collaboration}  % collaboration (optional)
%
%%%%%%%%%%%%%%%%%%%%%%%%%%%%%%%%%%%%%%%%%%%%%%%%%%%%%%%%%%%%%%%%%%%%%%%%%%%%%%%%%

%%%%%%%%%%%%%%%%%%%%%%%%%%%%%%%%%%%%%%%%%%%%%%%%%%%%%%%%%%%%%%%%%%%%%%%%%%%%%%%%%
% abstract
\abstract{%
There is evidence that the photoproduction of single and double pions off 
bound nucleons inside a nucleus are not only affected by Fermi motion but also
by other nuclear effects, such as final state interactions or meson rescattering.
We will present preliminary results of a high statistics measurement of single and 
double pion photoproduction of quasi-free protons and neutrons off the deuteron 
carried out at the Mainzer Microtron.
%Photoproduction of mesons offers an efficient way to determine the isospin 
%structure of the electromagnetic transition amplitudes 
%which are connected to the spin-flavor degrees of freedom of the nucleon.
%For this purpose, however, besides the reaction on the free proton at least 
%one reaction on the free neutron is mandatory. Due to the nonexistence of free 
%neutron targets, the only practicable way out is the investigation of the same 
%reaction on the deuteron and to isolate the desired information from the bound neutron 
%in quasi-free kinematics.
}
%
%%%%%%%%%%%%%%%%%%%%%%%%%%%%%%%%%%%%%%%%%%%%%%%%%%%%%%%%%%%%%%%%%%%%%%%%%%%%%%%%%

%%%%%%%%%%%%%%%%%%%%%%%%%%%%%%%%%%%%%%%%%%%%%%%%%%%%%%%%%%%%%%%%%%%%%%%%%%%%%%%%%
% main text
% for short contributions sections are optional
%\section{Introduction}
%  Meson photoproduction allows a detailed investigation of the excitation 
%spectrum of the nucleon and of the interactions of mesons with nucleons 
%and nuclei. Up to now much more attention has been paid to the electromagnetic 
%excitations of the free proton than to its neutron counter parts. 
%These are essential for the understanding of their isospin decomposition. 
%In addition in certain states the electromagnetic coupling of photons to 
%protons is rather different from that to neutrons. So far the sole experimental 
%possibility to investigate this subject is the quasi-free photoproduction of 
%mesons off neutrons bound in nuclei, in particular in the deuteron. As a 
%consequence the production cross section will of course be influenced by 
%nuclear Fermi motion and potentially also by nuclear final state 
%interaction effects (FSI). However, such effects can be studied by a 
%comparison of the free proton cross section to the quasi-free cross 
%section measured in coincidence with recoil protons. In the last few 
%years, the photoproduction of mesons off the deuteron were studied with 
%the Crystal Ball/TAPS setup at the MAMI accelerator in Mainz. We will report 
%some of the most interesting results of single and double $\pi^{0}$ photoproduction 
%off quasi-free protons and neutrons.

\section{Introduction}
The following paragraphs present preliminary results about single and 
double $\pi^{0}$ photoproduction off the deuteron that both originate from
the same experiment accomplished in December $2007$ at the Mainzer Microtron 
(MAMI) in Mainz, Germany. The MAMI electron beam facility produces a continuous 
photon beam with energies up to $1.5$ GeV. The photon beam was circularly polarized 
and the main detectors used in this experiments providing nearly full angular 
coverage are the \emph{Crystal Ball} calorimeter (CB) surrounding the target 
and the TAPS-detector which is placed as a forward 
wall. The separation of neutral and charged particles is done with plastic 
scintillators, either as bars arranged in a cylindrical setup surrounding 
the target (CB) or as hexagonally shaped vetos (TAPS). Furthermore a 
$\chi^{2}$-test is used to identify the photons stemming 
from the meson decay and to isolate the recoil neutron.\\
The single and double $\pi^{0}$ cross sections were measured throughout the second and third 
resonance region in coincidence with recoil protons (quasi-free exclusive 
reaction on the proton), in coincidence with recoil neutrons (quasi-free 
exclusive reaction on the neutron) and without a condition for the detection of recoil nucleons 
(quasi-free inclusive reaction). Both quasi-free exclusive reactions sum up 
to the quasi-free inclusive channel, since the contribution of the 
coherent process is negligible in the energy region of interest. 

\section{Results}
The left-hand side of figure \ref{XSsp} shows the preliminary single $\pi^{0}$ total cross 
section of the quasi-free exclusive reaction on the proton $\gamma+d\to\pi^{0}+p(n)$ 
(filled blue circles) together with the quasi-free exclusive reaction on the neutron 
$\gamma+d\to\pi^{0}+n(p)$ (filled red circles). The right-hand side of figure \ref{XSsp} 
shows the preliminary single $\pi^{0}$ total cross section of the quasi-free inclusive 
reaction $\gamma+d\to\pi^{0}+(N)$ (filled black circles) together with the sum of 
the two quasi-free exclusive reactions (open magenta circles). As well shown are 
previous results \cite{Kr99} for the inclusive reaction (open green circles) and the 
predicted cross sections from the theoretical models MAID \cite{MAID} (dashed lines) and 
SAID \cite{SAID} (full lines) folded with Fermi motion.

%%%%%%%%%%%%%%%%%%%%%%%%%%%%%%%%%%%%%%%%%%%%%%%%%%%%%%%%%%%%%%%%%%%%%%%%%%%%%%%%%
% the recommended way to include figures
\begin{figure}[htb]
\centering
%\subfloat[Quasi-free exclusive]
  \includegraphics[scale=0.25]{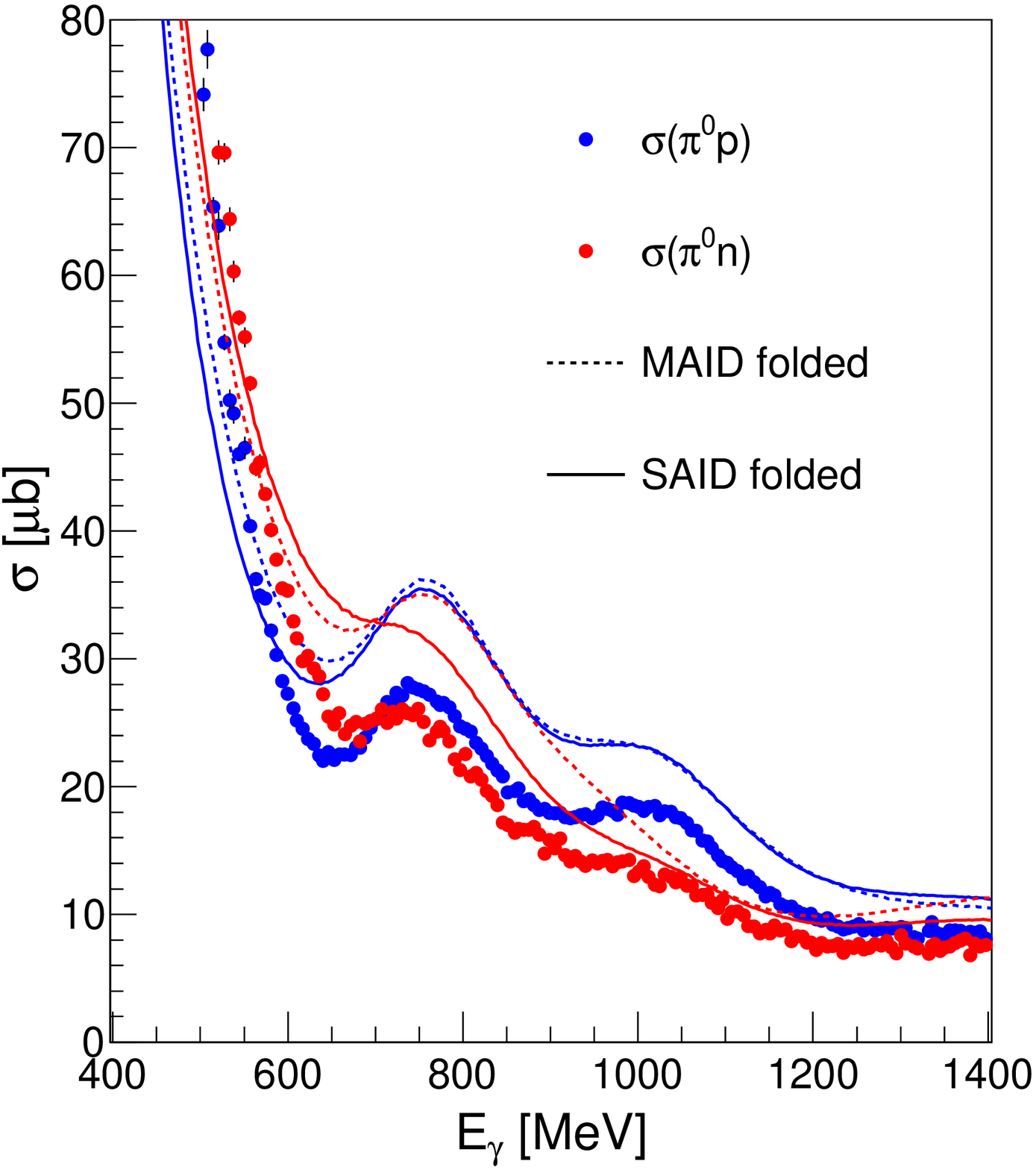}
%\subfloat[Quasi-free inclusive]
  \includegraphics[scale=0.25]{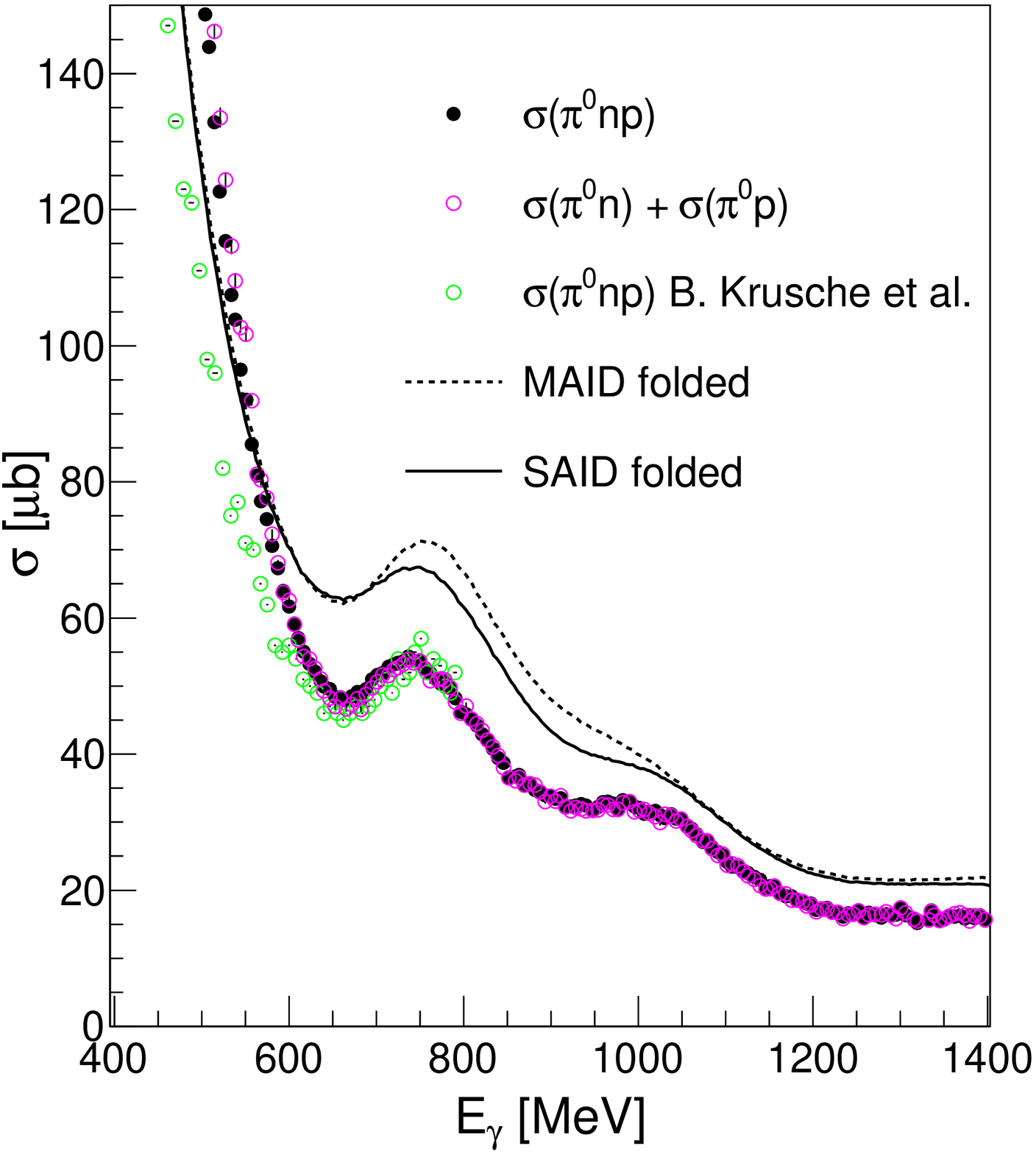}
\caption{Preliminary single $\pi^{0}$ total cross sections of the quasi-free exclusive 
(left-hand side) and quasi-free inclusive (right-hand side) reaction. The 
labelling is explained in the text.}\label{XSsp}
\end{figure}
%
%%%%%%%%%%%%%%%%%%%%%%%%%%%%%%%%%%%%%%%%%%%%%%%%%%%%%%%%%%%%%%%%%%%%%%%%%%%%%%%%%

It can be seen in figure \ref{XSsp} that the shapes of the measured cross sections 
(full red, blue and black circles) are in nice agreement with the theoretical MAID 
\cite{MAID} and SAID \cite{SAID} models but there is a disagreement in magnitude. 
Furthermore, the sum of the quasi-free exclusive cross sections (open magenta circles) 
add up perfectly to the measured quasi-free inclusive cross section (full black circles). 
This indicates a very clean identification of the reaction channels. 
In addition, the measured quasi-free inclusive cross section (full black circles) 
is in good agreement with earlier results from MAMI B \cite{Kr99}.\\
\newline
The left-hand side of figure \ref{dp} shows preliminary results for the $2\pi^{0}$ total cross sections of the quasi-free 
exclusive reaction on the proton $\gamma+d\to\pi^{0}\pi^{0}+p(n)$ (filled blue triangles)
together with the quasi-free exclusive reaction on the neutron $\gamma+d\to\pi^{0}\pi^{0}+n(p)$ 
(filled red triangles) and the quasi-free inclusive reaction $\gamma+d\to\pi^{0}\pi^{0}+(N)$ 
(filled black triangles). 

%\begin{figure}[htb]
%  \begin{center}
%    % please do not add file name extension this makes switching between latex and pdflatex easier
%    \includegraphics[width=0.5\textwidth]{CS_25_5_11.eps}
%    \caption{Measured preliminary double $\pi^{0}$ total cross sections. 
%The labelling is explained in the text.}
%    \label{XSdp}
%  \end{center}
%\end{figure}

The right-hand side of figure \ref{dp} illustrates the measured beam helicity asymmetries 
$I^{\odot} (\Phi)=1/P_{\gamma}(d\sigma^{+}-d\sigma^{-})/(d\sigma^{+}+d\sigma^{-})=
1/P_{\gamma}(N^{+}-N^{-})/(N^{+}+N^{-})$ \cite{Kr09} of the quasi-free exclusive 
reaction on the proton (full blue triangles) and on the neutron (full red triangles) for 
different regions of beam energy. The dashed lines correspond to a fit to the data. 

\begin{figure}[htb]
\centering
%\subfloat[Total cross sections]
  \includegraphics[width=0.4\textwidth]{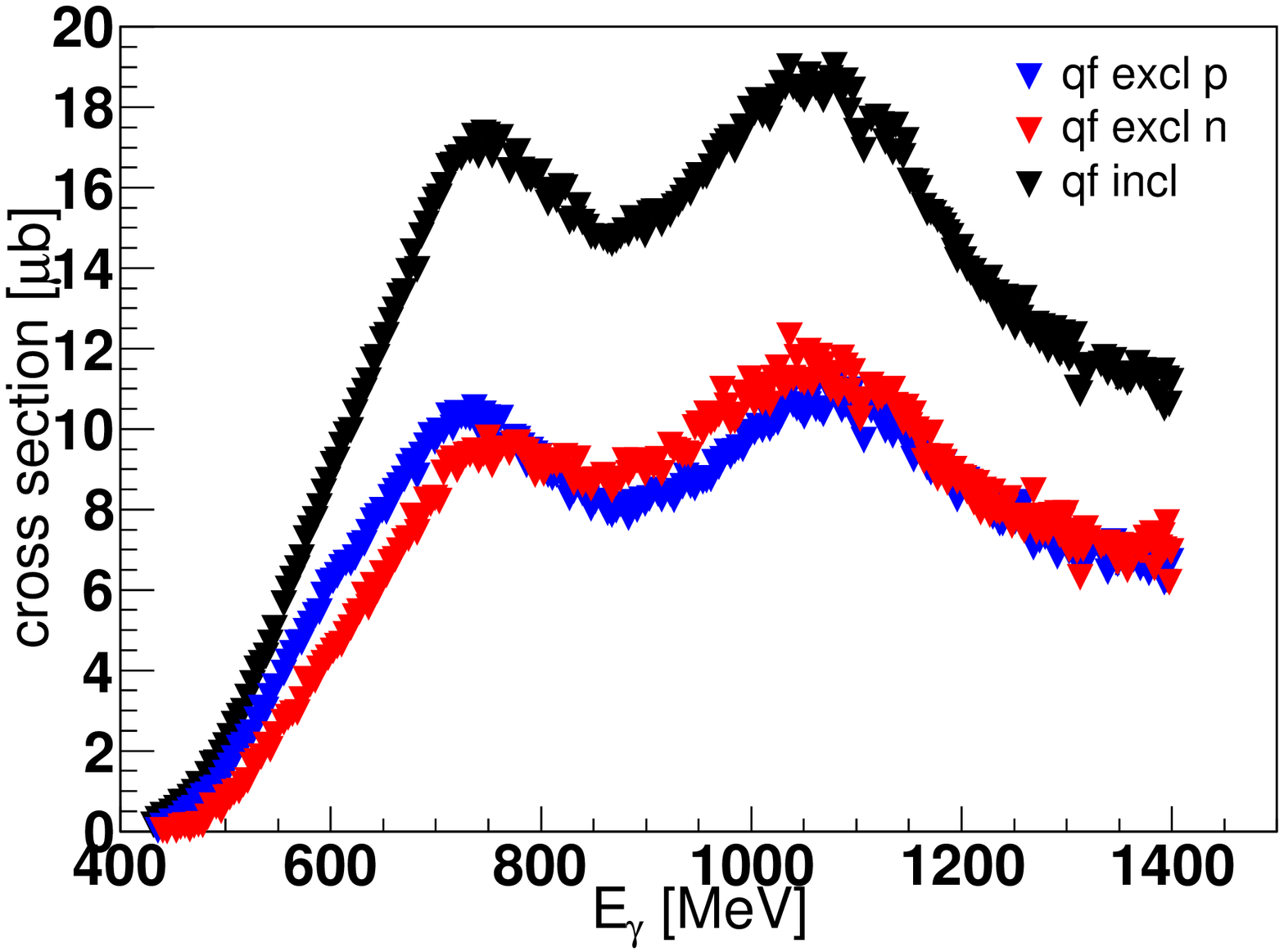}
%\subfloat[Beam helicity asymmetries]
  %{\includegraphics[width=0.4\textwidth]{I_Wr_pvsFreeP1Bin_cut2_new.eps}\label{HAdp2}}
%\caption{Measured preliminary beam helicity asymmetries. The 
%labelling is explained in the text.} 
  \includegraphics[width=0.5\textwidth]{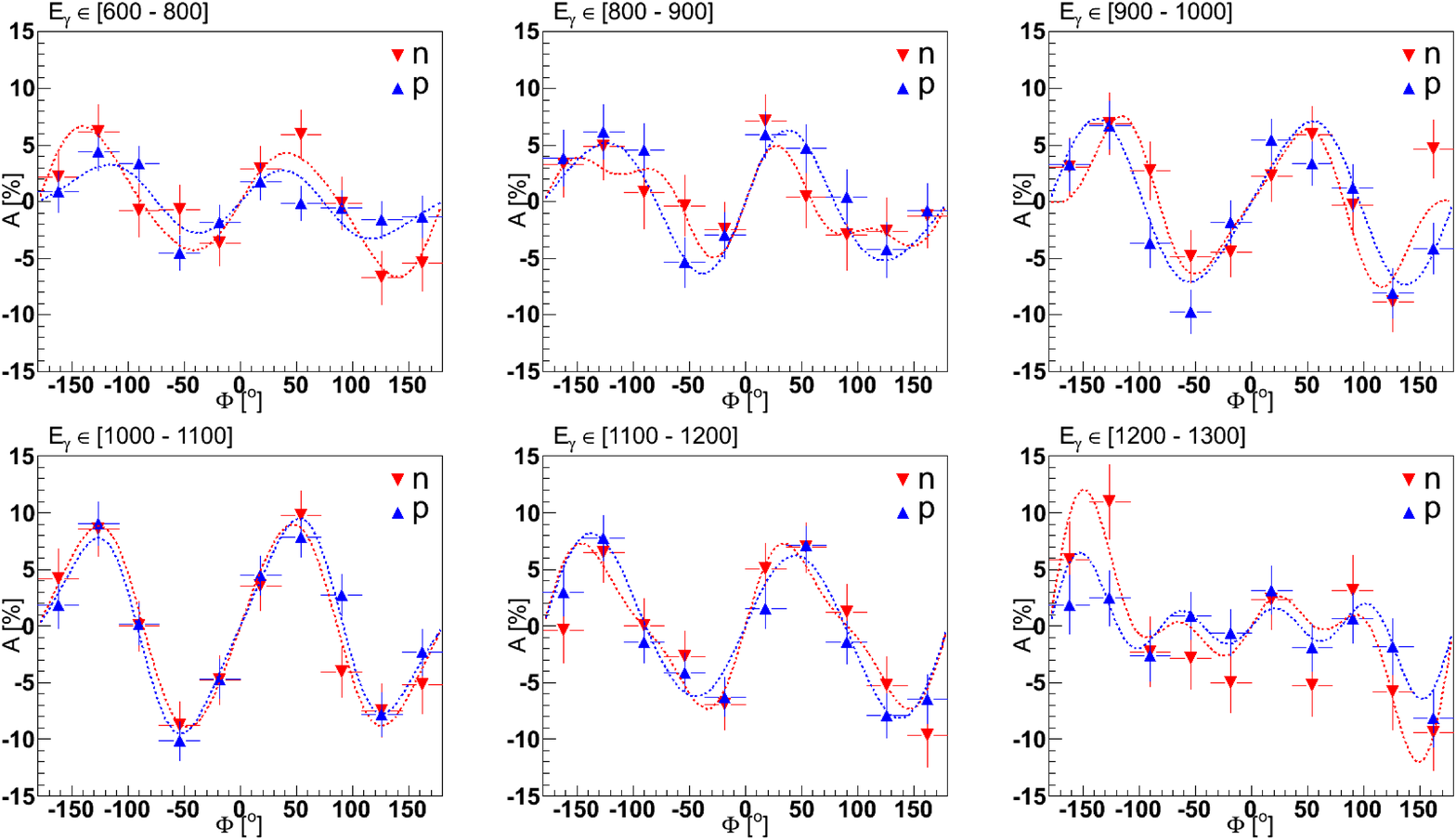}
\caption{Very preliminary double $\pi^{0}$ total cross sections (left-hand side) and beam helicity 
asymmetries (right-hand side). The labelling is explained in the text.}\label{dp}
\end{figure}

%%%%%%%%%%%%%%%%%%%%%%%%%%%%%%%%%%%%%%%%%%%%%%%%%%%%%%%%%%%%%%%%%%%%%%%%%%%%%%%%%
% the recommended way to include a LaTeX table
%\begin{table}[tb]
%  \begin{center}
%    \begin{tabular}{lccc}  
%      Patient        & Initial level($\mu$g/cc) & w. Magnet & w. Magnet and Sound \\
%      \hline
%      Guglielmo B.   &                    0.12  &      0.10 &               0.001 \\
%      Ferrando di N. &                    0.15  &      0.11 &          $< 0.0005$ \\
%    \end{tabular}
%    \caption{Blood cyanide levels for the two patients.}
%    \label{tab:blood}
%  \end{center}
%\end{table}
%
%%%%%%%%%%%%%%%%%%%%%%%%%%%%%%%%%%%%%%%%%%%%%%%%%%%%%%%%%%%%%%%%%%%%%%%%%%%%%%%%%

\section{Interpretation}

The identical shape of the measured single $\pi^{0}$ total cross sections 
and the theoretical MAID \cite{MAID} and SAID \cite{SAID} models (see figure \ref{XSsp}) demonstrates 
a correct understanding of the resonance contributions to the different reactions, i.e. mainly 
$D_{13}(1520)$, $F_{15}(1680)$ to single $\pi^{0}$ photoproduction on the proton and 
$D_{13}(1520)$, $D_{15}(1675)$ on the neutron. The discrepancy in absolute height between the measured 
single $\pi^{0}$ total cross sections and the theoretical models (see figure \ref{XSsp}) can not 
be explained by nuclear Fermi motion. The fact that folding the theoretical models with Fermi motion 
does not overcome this problem reveals the importance of other nuclear effects, such as final 
state interactions, meson rescattering or others. This discrepancy was already earlier observed 
by B. Krusche \etal \cite{Kr99} and H. Shimizu \cite{Sh09}. H. Shimizu reported an overestimation of the data by models of $\sim125\%$, as shown at the left-hand side of figure 
\ref{Scale}. The same level of overestimation was observed in this work, as depicted on the 
right-hand side in figure \ref{Scale}, where the models are scaled down by a factor of $0.8$.

\begin{figure}[htb]
\centering
%\subfloat[H. Shimizu \cite{Sh09}]
  \includegraphics[width=0.35\textwidth]{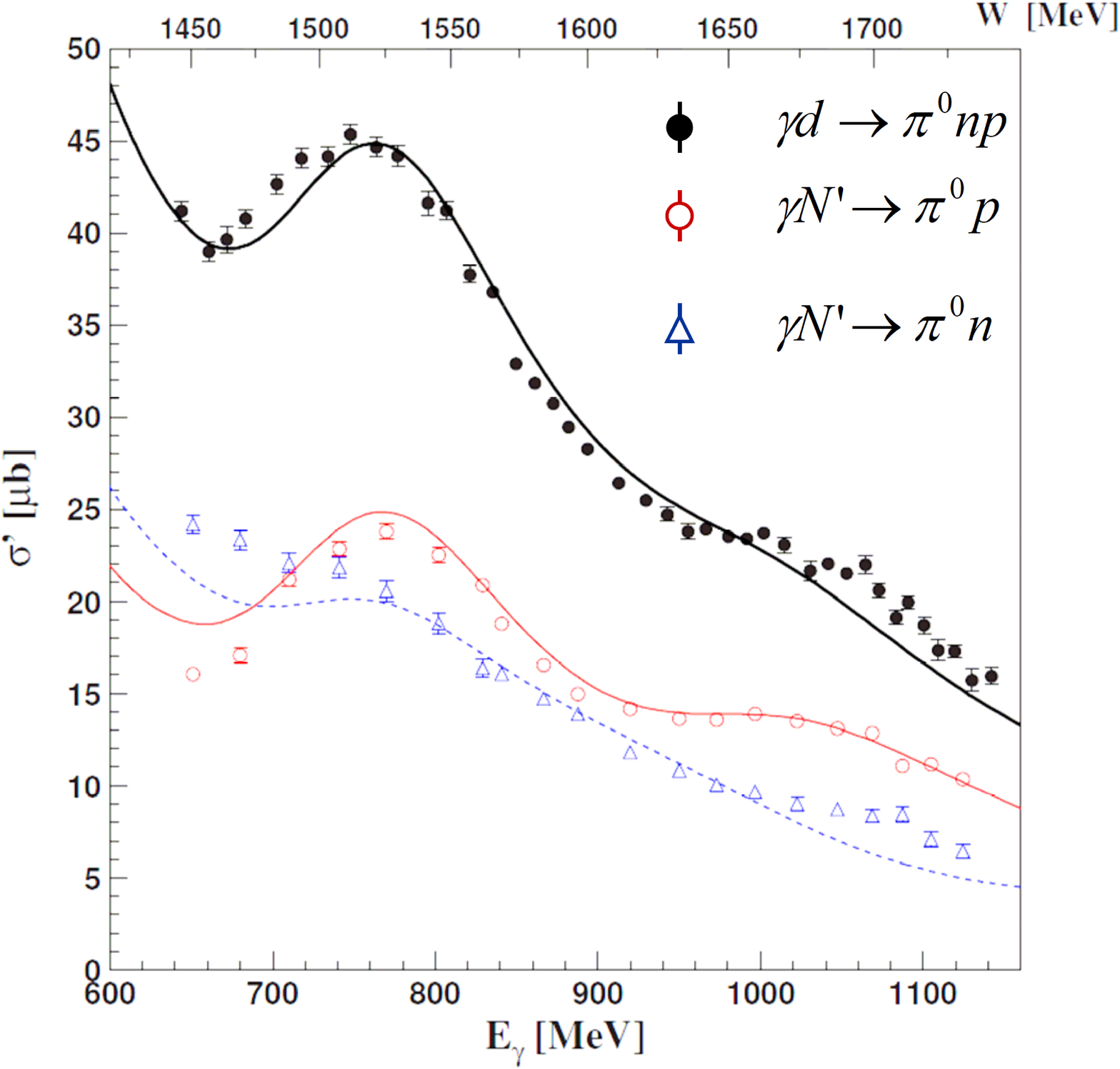}
%\subfloat[This work]
  \includegraphics[width=0.4\textwidth]{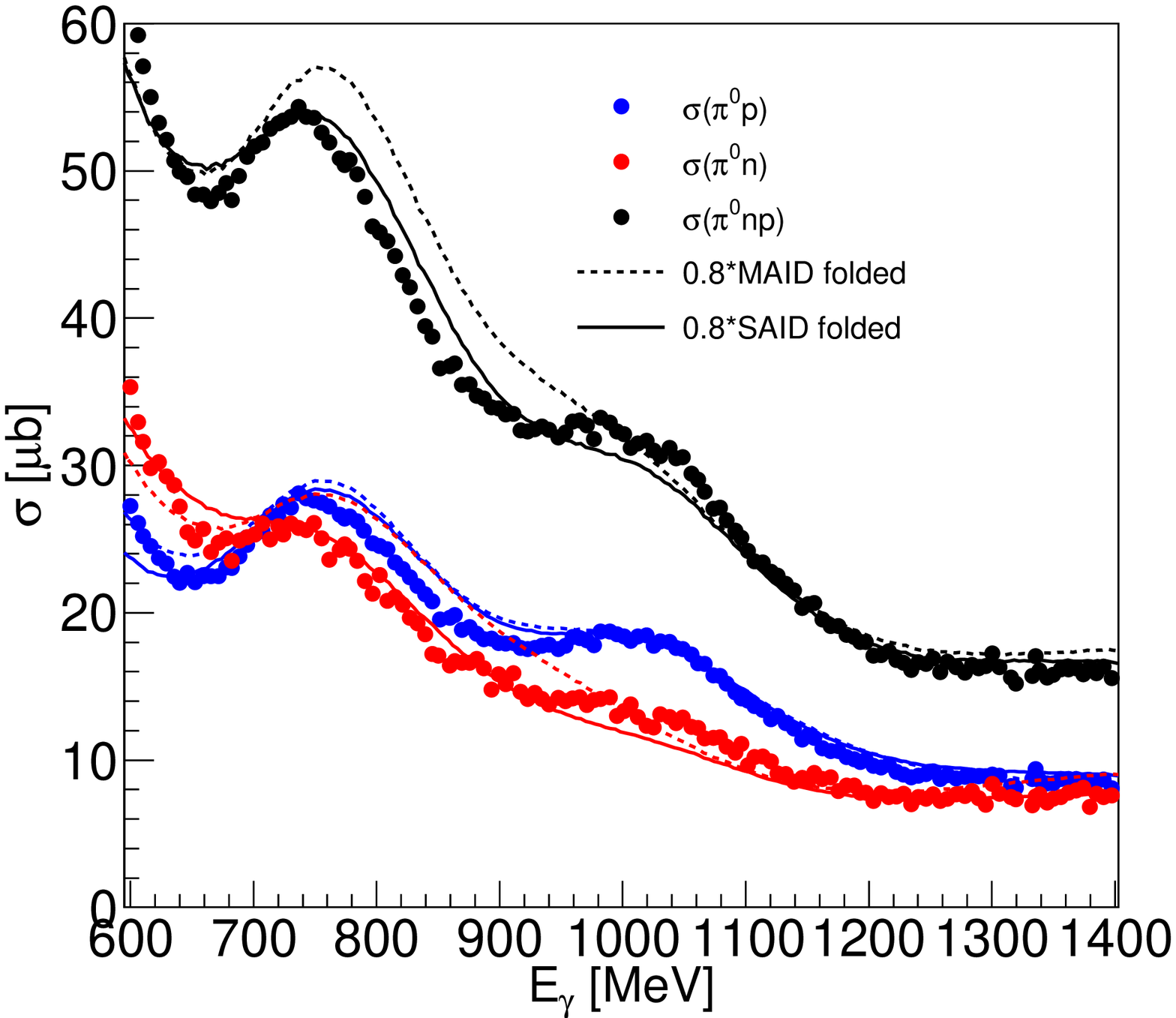}
\caption{Quasi-free exclusive total cross sections compared to the MAID \cite{MAID} and SAID \cite{SAID} 
models scaled down by a factor $0.8$. Left-hand side: H. Shimizu \cite{Sh09}, right-hand side: This work. 
Note the reverted color coding on the left figure between red and blue.} 
\label{Scale}
\end{figure}

Double $\pi^{0}$ photoproduction is mainly used to study the properties of sequential decays since this 
is the dominated decay mechanism. Even though the MAID model \cite{MAID} predicts a nearly identical total 
cross section for the production on the proton and on the neutron, the resoncance contribution to the two 
reactions is rather different. For example, the electromagnetic excitation of the $F_{15}(1680)$ $(D_{15}(1675))$ is 
predicted to be much stronger on the proton (neutron) than on the neutron (proton). For this reason one would 
suggest that sensitive quantities such as the beam helicity asymmetry would depend on such a coupling and hence 
will not be the same on the proton and on the neutron.\\
The measured $2\pi^{0}$ total cross sections on the proton and neutron (left-hand side of figure \ref{dp}) show 
no big difference and hence confirm the predictions. However, it is rather astonishing that 
the measured beam helicity asymmetries for the proton and neutron are as well identical within the error bars. 
Additionally, the results on the quasi-free proton were defolded from Fermi motion and compared to published data on the 
free proton and are in good agreement with each other (not shown here) which indicates a correct 
reconstruction of the reaction. Yet, earlier works \cite{Kr09} also contradicted many model predictions 
concerning the asymmetries, therefore further input is needed in order to understand this behavior.

%The results from double $\pi^{0}$ photoproduction are rather contradicting to any theoretical predictions. 
%The work done by M. Oberle et al. and earlier published results \cite{Ze11} show nearly the same 
%beam helicity asymmetries for both the proton and the neutron whereas theory predicts a different 
%electromagnetic excitation of the resonances for the proton than for the neutron. For example the $F_{15}$ ($D_{15}$) 
%couples much stronger (weaker) to the proton than to the neutron. Furthermore most of the theoretical models 
%are not able to reproduce the measured asymmetries at all. 

\acknowledgements{
   This work is supported by Schweizerischer Nationalfonds (SNF), Deutsche Forschungsgemeinschaft (DFG) and EU/FP6. 
All the work on the double $\pi^{0}$ channel is done by M. Oberle \etal.
}
%%%%%%%%%%%%%%%%%%%%%%%%%%%%%%%%%%%%%%%%%%%%%%%%%%%%%%%%%%%%%%%%%%%%%%%%%%%%%%%%%
% bibliographic items can be constructed using the LaTeX format in SPIRES
% see http://www.slac.stanford.edu/spires/hep/latex.html
% SPIRES will also supply the CITATION line information; please include it

%
%%%%%%%%%%%%%%%%%%%%%%%%%%%%%%%%%%%%%%%%%%%%%%%%%%%%%%%%%%%%%%%%%%%%%%%%%%%%%%%%%

}  % do not remove

%%% Local Variables: 
%%% mode: latex
%%% TeX-master: "../hadron2011.tex"
%%% End: 


\begin{thebibliography}{99}
  
\bibitem{Kr99}
  B. Krusche \etal, Eur. Phys. J. A 6 (1999) 309
\bibitem{MAID}
  D. Drechsel, O. Hanstein, S.S. Kamalov and L. Tiator, Nucl. Phys. A \textbf{645} (1999) 145
\bibitem{SAID}
  R. Arndt \etal, VPI and SU Scattering Analysis Interactive Dialin
\bibitem{Kr09}
  D. Krambrich, F. Zehr, A. Fix, L. Roca \etal, Phys. Rev. Lett. 103 (2009) 052002
\bibitem{Sh09}
  H. Shimizu, Slides presented at the NNR workshop 2009, June 8.-10., 2009, Edinburgh
\end{thebibliography}
\end{document}